Vladislav G. Polnikov

# INTEGRATED MODEL
# FOR A WAVE BOUNDARY LAYER

A.M. Obukhov Institute of Atmospheric Physics of Russian Academy of Sciences

Pyzhevskii lane 3, 119017, Moscow, Russia**,**

E-mail: polnikov@mail.ru

Tel: +7-916-3376728

Fax: +7-495-9531652

## Abstract

In the paper a new version of semi-phenomenological model is constructed, which allows to calculate the friction velocity $u_*$ via the spectrum of waves $S$ and the wind at the standard horizon **W**. The model is based on the balance equation for the momentum flux, averaged over the wave-field ensemble, which takes place in the wave-zone located between troughs and crests of waves. Derivation of the balance equation is presented, and the following main features of the model are formulated. First, the total momentum flux includes only two physically different types of components: the "wave" part $\tau_w$ associated with the energy transfer to waves, and the "tangential" part $\tau_t$ that does not provide such transfer. Second, component $\tau_w$ is split into two constituents having different mathematical representation: (a) for the low-frequency (energy-containing) part of the wave spectrum, the analytical expression of momentum flux $\tau_w$ is given directly via the local wind at the standard horizon, **W**; (b) for the high-frequency part of the wave spectrum, flux $\tau_w$ is determined by friction velocity $u_*$. Third, the tangential component of the momentum flux is parameterized by using the similarity theory, assuming that the wave-zone is an analogue of the traditional friction layer, and in this zone the constant eddy viscosity is realized, inherent to the wave state. The constructed model was verified on the basis of simultaneous measurements of two-dimensional wave spectrum $S$ and friction velocity $u_*$, done for a series of fixed values of *W*. It is shown that the mean value of the relative error for the drag coefficient, obtained with the proposed model, is 15-20% , only.

## 1. Introduction

This work is devoted to constructing a conceptually new model of the dynamic boundary layer of the atmosphere (or the wave boundary layer, WBL, in terms of Chalikov (1995) ) and is a natural extension of our previous paper (Polnikov 2009). Herewith, it is believed that the main purpose of the WBL-model is to establish the mathematical relation between characteristics of the boundary layer of the atmosphere (in particular, the friction velocity $u_*$) with the two-dimensional spatial spectrum of wind waves $S(k_x, k_y)$ (or the frequency-angular analogue, $S(\omega,\theta)$), and the mean wind at standard horizon **W**. The mathematical basis of this model is the balance equation for the momentum flux of the form[1]

$$\tau_{\mathrm{mod}}(W, u_*, S) = u_*^2 . \qquad (1)$$

The physical content of the left-hand side of (1) must follow from the fundamental equations of hydrodynamics.

The task of constructing a physical model of the WBL can be solved in different ways. In Polnikov (2009) a classification of approaches to solving this problem is given where semi-phenomenological models (Janssen 1991; Zaslavskii 1995; Chalikov 1995; Makin and Kudryavtsev 1999) were mentioned (among numerous others) as the most acceptable in practical terms. More recent alternatives (Kudryavtsev and Makin 2001; Makin and Kudryavtsev 2002) also belong to this class[2]. In the terms mentioned, the advantages of such models are their physical justification based on the equations of hydrodynamics and the lack of overly cumbersome analytical or numerical details. In practical models the latter are replaced by a number of quite clear and simple phenomenological constructions. Based on the analysis mentioned, the models by Zaslavskii (1995) and Makin and Kudryavtsev (1999) were indicated as the most promising ones for further development. The disadvantages of these models were mentioned in Polnikov (2009) as well. Taking in mind these findings, we will try to advance further in the problem.

---

[1] Here and hereafter, the momentum flux is normalized on the air density $\rho_a$.

[2] These models were not analyzed in Polnikov (2009) due to technical reasons.

The physical basis of all these WBL-models is one or another version of the analytical representation of the left-hand side of balance equation (1) by a set of summands which generally has the form (Makin and Kudryavtsev 1999)

$$\tau_w + \tau_v + \tau_t = \tau = u_*^2 . \qquad (2)$$

In the left-hand side of (2) it is used to distinguish three components of the total stress $\tau$ : the wave component $\tau_w$, viscous one $\tau_v$, and turbulent one $\tau_t$. The first of them is usually associated with the transfer of energy from the wind to waves, whilst the last two create the tangential stress at the interface, not affecting the energy of waves. The problem is to find an adequate analytical representation for each of these components[3].

In semi-phenomenological approach it is widely accepted (see the same references) to describe the wave part of momentum flux $\tau_w$ via the known function $IN(S,\mathbf{W})$ that corresponds to the rate of energy pumping waves by the wind. It is traditionally represented as (Komen et al. 1994; Polnikov 2009)

$$IN(S,\mathbf{W}) = \beta(\omega,\theta,W,\theta_w)\omega S(\omega,\theta) , \qquad (3)$$

where $\beta(\omega,\theta,W,\theta_w)$ is the wave-growth increment usually accepted as a semi-empirical dimensionless function of their parameters. In this case, at the conditional mean surface level z = $<\eta(\mathbf{x},t)>$ = 0, one can write

$$\tau_w(z=0) \equiv \tau_w(0) = \rho_w g \int \frac{k\cos(\theta)}{\omega} IN\left(S,\mathbf{W},\omega,\theta\right) d\omega d\theta . \qquad (4)$$

Without dwelling on the technical point of choosing function $\beta(\omega,\theta,W,\theta_w)$, one can assume that the mathematical representation of flux $\tau_w$ is known, and its estimation does not cause any principal difficulties.

[3] It should be especially noted that the equation of the form (2) is traditionally ascribed to a narrow atmospheric layer located directly at the air-water interface (see the mentioned references). In this case, it is implicitly assumed that the flux to a random and non-stationary surface is essentially similar to the flux towards a solid surface. The stochasticity of a wave field, implying the presence of abrupt changes in shape of the air-water interface, and its non-stationarity due to multiple vertical wave motions, are not taken into account in such approach.

Regarding to the procedure of calculating values $\tau_t$ and $\tau_v$, in contrast to estimation of $\tau_w$, there is not a unity of approach. Therefore, in each of the above-mentioned versions for WBL-models, calculations of these quantities vary considerably.

For example, in the model (Makin and Kudryavtsev 1999), the calculation of $\tau_t$ is based on the well-known theoretical relation for the turbulent part of the momentum flux (Monin and Yaglom 1971)

$$\tau_t(z) = K\frac{\partial W(z)}{\partial z}, \tag{5}$$

where the dimensional coefficient *K* has a meaning of the turbulent viscosity in the WBL. Using representation (2) (considered beyond the space of influence of $\tau_v$), by constructing an additional specific model for the turbulent viscosity in the WBL of the atmosphere, Makin and Kudryavtsev (1999) have obtained

$$W(z) = u_*^2 \int_{z_0^v}^{z} \left[1 - \frac{\tau_w(z)}{u_*^2}\right] K^{-1} dz = \frac{u_*}{\kappa} \int_{z_0^v}^{z} \left[1 - \frac{\tau_w(z)}{u_*^2}\right]^{3/4} d(\ln z) \equiv \frac{u_*}{\kappa} F(z). \tag{6}$$

Result (6) allows to complete the solution by using a known expression for the vertical profile of the wave flux $\tau_w(z)$. In the referred paper, the latter is adopted as the exponentially decaying function: $\tau_w(z) = \tau_w(0)\exp(-z / L(k))$ (see details in the references). Herewith, the essential point of the model by Makin and Kudryavtsev (1999) is the assertion that just the turbulent component of friction velocity, $u_{*_t} = (\tau_t)^{1/2}$, is to be taken in the integrand of (4) (instead of the full friction velocity, $u_* = (\tau)^{1/2}$), in the course of calculating $\tau_w$ by formula (4). However, the results of the model verification, obtained in (Polnikov 2009), show that this approach to constructing the WBL-model does not provide an adequate range of variability for dependence $u_*(S,W)$, and requires some modification.

The sequential works by the same authors (Kudryavtsev and Makin 2001; Makin and Kudryavtsev 2002) serve as an example of such, rather significant modification of the approach mentioned. Since they have not been discussed previously in (Polnikov 2009), it is worthwhile to bring shortly the following important points. Thus, Kudryavtsev and Makin (2001) rejected the idea of using the differential relation (5), and began to assess the value of $\tau_t$ by accepting the idea of matching the linear profile of the average wind speed $W(z)$ with the standard logarithmic velocity profile at the boundary of the molecular viscous sublayer (where profile $W(z)$ depends on $\tau_t$). It means that they consider equation (2) at the interface (see footnote 3).[4]

Moreover, in order to enhance the dynamics of variability of the final solution $u_*(S,W)$, the authors formulated the idea of appearing the additional stress $\tau_{fs}$ caused by the air-flow separation due to breaking crests of the high-frequency components of a wave spectrum. Herewith, they have assumed that the contribution of $\tau_{fs}$ to $\tau_w$ is the essential complement to the traditional representation of $\tau_w$, which can not be compensated by an appropriate choice of the of growth increment $\beta(\omega,\theta,W,\theta_w)$. The next paper (Makin and Kudryavtsev 2002) is the development of this approach for the case of the dominant-wave breaking.

Ultimately, the equation for $u_*(S,W)$ takes the form (Kudryavtsev and Makin 2001)

$$\iint_{k,\theta} \tilde{\beta}(...)B(k,\theta)\cos\theta d\theta d\ln k + C_{fs} \iint_{k<k_{fs},\theta} \frac{\tilde{\beta}(...)u_*^2}{c_{ph}^2(k)} B(k,\theta)\cos^3\theta d\theta d\ln k + C_t \ln[\frac{c_\nu \nu}{u_* z_0(W)}] = 1 \quad (7)$$

where the first term corresponds to the traditional flux $\tau_w(u_*,S)$ expressed in terms of the saturation spectrum $B(k,\theta) = k^3 S(k,\theta)$; the second term is $\tau_{fs}(u_*,S)$; and the third one corresponds to the term $\tau_t$, in which the viscosity of air $\nu$ and the roughness height $z_0(W)$ are inevitable parameters of the model (for detailed explanation, see the original).

In Kudryavtsev and Makin (2001) it was shown that model (7) reproduces quite well numerous observed dependences of different WBL-parameters on wave characteristics. Moreover, in (Babanin and Makin 2009), by attracting simultaneous measurements of the wave spectrum $S(\omega,\theta)$ and magnitude of $u_*$, it was shown that for the model discussed, the mean relative error of the total flux (or the drag coefficient $C_d$), given by the ratio

$$\rho_{Cd} = (u_{*mod} / u_*)^2 - 1 \ , \qquad (8)$$

is of the order of 20-25% in a wide range of values for $u_*$. All the said testifies to the significant advantage of the new model (Kudryavtsev and Makin 2001) compared with the earlier one (Makin and Kudryavtsev 1999).

[4] Hereafter, such approach is suitable to be called “integrated”, and the proper model is called integrated, in contrast to the approach based on ratios (5) and (6).

However, peering and detailed consideration of the concept of constructing the integrated model (Kudryavtsev and Makin, 2001) raises several questions.

First. The introduction of term $\tau_{fs}$ is interesting itself from the standpoint of physics, though it does somewhat artificially complicate the task of constructing the WBL-model. Indeed, due to linearity in the spectrum, the main contribution of the air-flow separation stress $\tau_{fs}$ to the total stress $\tau$, in fact, can be accounted for in the traditional representation for $\tau_w$ by choosing proper parameters for the empirical increment $\beta(\omega,\theta,W,\theta_w)$. Appropriateness of the said is provided by the fact that inevitable breaking events are automatically taken into account in the empirical parameterization of $\beta(\omega,\theta,W,\theta_w)$ (Drennan et al. 1999; Komen at al. 1994; Chalikov and Rainchik 2010; among others).

In addition, after averaging the balance equation over the wave-field ensemble (see details in Section 2), a possible contribution of $\tau_{fs}$ to the tangential stress becomes uncertain, and this circumstance should be taken into account in the assessment of $\tau_t$. Thus, the introduction into consideration the air-flow separation, as well as the introduction of breaking events, complicates the basing a validity of using the traditional (molecular) viscous sublayer for the determination of tangential stress $\tau_t$, realized in (Kudryavtsev and Makin 2001).

Second. The validity of using the traditional viscous sublayer in the situation with a random, highly non-stationary, and spatial-inhomogeneous surface experiencing the breaking, invokes a serious doubt in the method of assessing $\tau_t$ used in the model said. Breaking and randomness of the interface are clearly not in accordance with the traditional approach based on existence of molecular viscous sublayer, applicable to a firm and fixed surface. It seems that in view of stochasticity of the interface, the traditional concept of the viscous sublayer should be refused and properly replaced.

Third. The final representation of the balance equation in form (7), resulting from introduction a set of postulates, hypotheses, and fitting constants of the model ($C_{fs}, k_{fs}$ $C_t, c_v$), is very cumbersome and difficult to treat it due to its irrational kind with respect to $u_*$. Therefore, there arises a natural need in constructing a simpler, but equally physically meaningful WBL-model.

Thus, all the said above is the basis for attempts to construct a new semi-phenomenological WBL-model in a frame of less complicated and physically reasonable assumptions and fitting parameters. This paper is aimed to solve this problem.

The structure of the paper is the following. In Section 2 we mention shortly the main points of the balance equation derivation with the aim to introduce the new concept of “the wave-zone”. That allows us to get new treating the terms of this equation. In Sections 3 and 4 parameterizations for two constituents of wave-part stress $\tau_w$ are specified. Section 5 is devoted to constructing parameterization for $\tau_t$ by means of the similarity theory. The method and results of the model verification are given in Section 6, and the final conclusive remarks are presented in Section 7.

## 2. Derivation of the Balance Equation. Introducing the Concept of "the Wave-zone"

In order to establish the physical content of the left-hand side of equation (1) applied to the problem with a random wavy interface, and taking into account the comments made in footnote 3, we briefly consider the balance equation derivation for a wavy surface (the main calculations in this section are a courtesy of VN Kudryavtsev). We show that the kind of representation of the left-hand side of (1) is largely determined not by the detailed account of the dynamic equations at the wavy interface $\eta(\mathbf{x},t)$ but is done by the rules of averaging the balance equation.

To obtain equation (1), the Navier-Stokes equations are used as initial ones, written (for generality) for the three-dimensional, unsteady turbulent flow of air with the mean speed profile $W(z)$ over the wavy interface:

$$\partial u_\alpha / \partial t + \partial (u_\alpha u_i) / \partial x_i = -\partial p / \partial x_\alpha + \partial \tau_{\alpha i} / \partial x_i . \quad (9)$$

Here, for index $\alpha(=1,2)$ it is sufficient to take value 1, whilst the index *i* takes the values 1,2,3, corresponding to *x, y, z* components of the velocity field **u**; the pressure field *p* is given in the normalization on the density of air; $\tau_{\alpha i}$ is the viscous-stress tensor; the remaining notations are taken from the monograph (Monin and Yaglom 1971). The task is to obtain equations for the momentum fluxes. For this purpose a procedure of integrating equation (9) over the vertical variable from the interface $\eta(\mathbf{x},t)$ to horizon *z* located far from the wavy surface is used. In such a case, the formula of differentiation of the integral by parameter (Leibniz's formula) is applied (Monin and Yaglom 1971):

$$\frac{\partial}{\partial p}\left(\int_{a(p)}^{b(p)} f(x,p)dx\right) = \int_a^b \frac{\partial}{\partial p} f(x,p)dx + \left[ f(x,p)\frac{\partial b(p)}{\partial p}\right]\Big|_{x=b} - \left[ f(x,p)\frac{\partial a(p)}{\partial p}\right]\Big|_{x=a} . \quad (10)$$

After integrating (9) and using (10) with $a = \eta(\mathbf{x},t)$ and $b = z$, it follows (Makin et al. 1995)

$$\frac{\partial}{\partial t}\left(\int_\eta^z u_\alpha dx_3\right) + \frac{\partial}{\partial x_\beta}\left(\int_\eta^z u_\alpha u_\beta dx_3\right) - u_\alpha u_3\big|_z = \frac{\partial}{\partial x_\alpha}\left(\int_\eta^z p dx_3\right) + \frac{\partial}{\partial x_\beta}\left(\int_\eta^z \tau_{\alpha\beta} dx_3\right) + \left(p\frac{\partial \eta}{\partial x_\alpha}\right)\Big|_\eta - \tau_{\alpha 3}\big|_z + (\tau_{\alpha 3} + \tau_{\alpha\beta}\frac{\partial \eta}{\partial x_\beta})\Big|_\eta \quad (11)$$

where the appearing index $\beta$ is analogous to index $\alpha$.

Now, it is necessary to perform the spatial and temporal averaging (11) (for $\alpha = 1$), supposing for simplicity that $u_1(x,z) = W(z) + u'(x,z)$, $u_2(x,z)$=0, $u_3(x,z) = w'(x,z)$ (where the prime means the turbulent fluctuations of the air velocity). Herewith, it should be noted that the averaging is to be done both on the turbulent scales and the scales of the wave-field variability.

The first type of averaging leaves the integral forms in (11) practically unchanged, while the "free" terms in (11) yield the momentum fluxes that we are looking for.

For a steady flow over a solid and horizontal surface, the integral terms in (11) vanish. For a moving and horizontally inhomogeneous surface, these terms can also be cleared but only if one passes to the scales of horizontally homogeneous and steady-state description of the wave field, i.e. after the wave-ensemble averaging. Incidentally, it is the averaging that allows us to pass to the spectral representation for waves, used in the WBL-model representation. As a result of these actions, the final equation of the form (1) appears, which can be written in the form

$$\left\langle (< p\partial\eta / \partial x >) + < \tau_x > \right\rangle = \left\langle - < u_x u_z > \right\rangle \tag{12}$$

where the small angle brackets <..> denotes averaging over the turbulent scales, and large ones does the wave-ensemble averaging. Note that balance equation (12) is not ascribed to the certain wave profile, as well as the wave spectra are not ascribed to the certain water level. In this case, generally speaking, it should be formulated a special treatment of the terms of equation (12), due to non-stationarity and inhomogeneity of interface $\eta(x,t)$.

Traditionally, the first term in the left-hand side of (12) is interpreted as the momentum flux associated with the work of pressure forces, while the second term to the viscous flux component. It means that the meaning of $\tau_w$ is prescribed for the first term, and the meaning of ($\tau_\nu + \tau_t$) does for the last one (see eq. (2)). Herewith, equation (12) is often interpreted as the balance equation written at the interface (see footnote 3 and the treatment of viscous terms in Kudryavtsev and Makin (2001)), implying that the average over the wave scales is realized by sliding along the current border between the media (Figure 1a).

Though, such kind averaging does implicitly assume the horizontal and temporal invariance of the vertical structure of the WBL (i.e. uniformity in the mean-wind profile $W(z)$ for any value of z, measured from the current position of the boundary), regardless to the wave phase. It is caused by wishing to use directly the traditional interpretation of the terms, mentioned above.

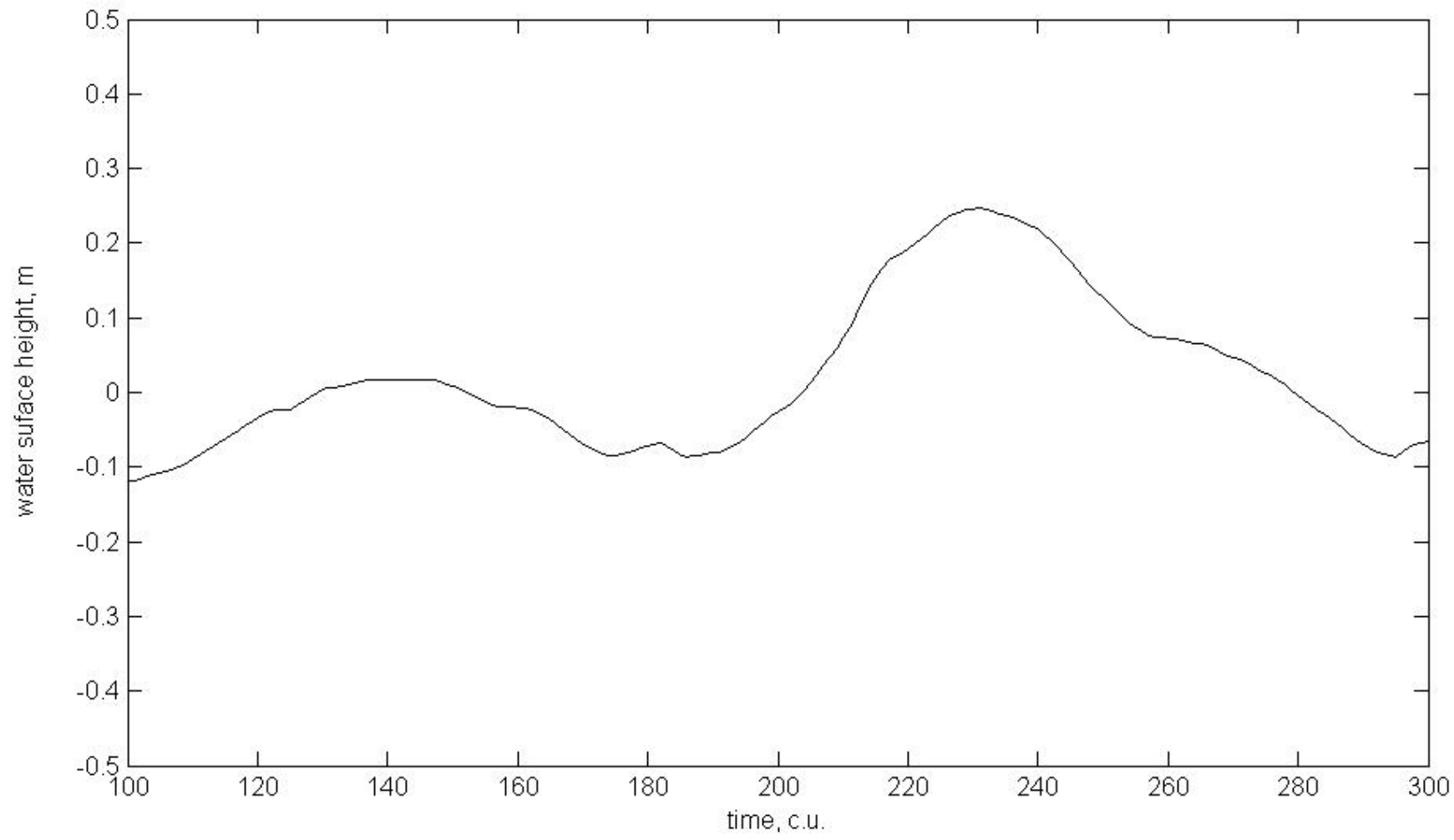

a)

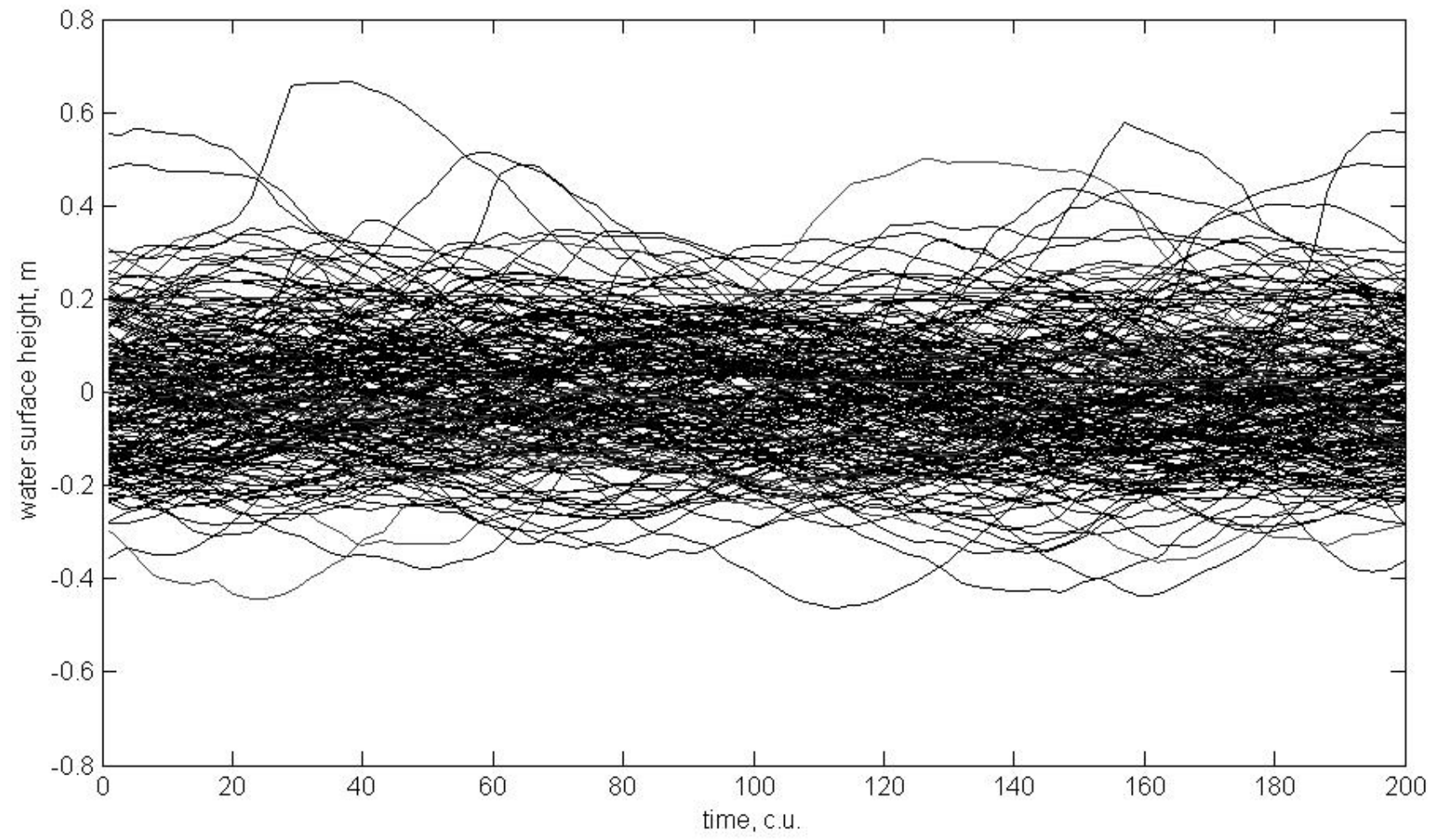

b)

Fig. 1. a) a part of wave record $\eta(x,t)$, b) an ensemble of two hundreds parts of the same wave record $\eta(x,t)$. The time scales are given in conventional units.

However, it is easy to imagine that in this, "monitor" coordinate system, the conditions of stationarity and horizontal homogeneity of the mean motion are violated (to say nothing about breaking), despite of the necessity of existing these conditions for removing the integral terms in (11).

In fact, the dynamics of an air flow over a wavy surface (and the vertical structure of the WBL) varies significantly on the ridges, troughs, windward and leeward sides of the wave profile, i.e. the horizontal and temporal invariance of the vertical structure of WBL is violated.

This "speculative", but fairly obvious conclusion is confirmed by the numerical solution of the Navier-Stokes equations (see, for example, Li et al. 2000; Sullivan et al. 2000; Chalikov and Rainchik 2010). Therefore, the result of averaging equation (11) for the case of a random wave surface, which is required not only to eliminate the non-stationarity and horizontal inhomogeneity of the vertical structure of WBL, but also for the transition to a spectral representation of waves, gives rise to search for different interpretation of (12) .

To achieve this goal, it is more natural to treat the averaging equation (11) as the averaging performed over the statistical ensemble of wave surfaces, conventionally depicted in Fig. 1b. Clearly, in this case, that equation (11) is averaged over the entire “wave-zone” located between the certain levels of troughs and crests of waves. This zone occupies the space from -*H* to *H* in vertical coordinate measured from the mean surface level, and the value of *H* is of the order of the standard deviation *D* given by the ratio

$$D = \sqrt{\left\langle \left( \eta - \langle \eta(\mathbf{x},t) \rangle \right)^2 \right\rangle} . \qquad (13)$$

Besides to the said, in the course of ensemble averaging, each term in the left-hand side of (12) undergoes to such changes which can not be identified and treated in advance, i.e. their contributions to $\tau_w$ и $\tau_t$ can not be distinguished. Therefore, we can assume that in the left-hand side of the final balance equation there are only two types of terms: the generalized wave stress $\tau_w$ (traditionally called “the form drag”) which corresponds to the entire transfer of energy from wind to waves; and the generalized tangential stress (called “the skin drag”) which is not associated with the wave energy. Thus, the resulting balance equation becomes somewhat "simpler" and takes the form

$$\tau_w(W, u_*, S, \omega_p) + \tau_t(W, u_*, S, \omega_p) = \tau_{tot} . \qquad (14)$$

In (14) contributions of the individual terms of the left-hand side of equation (12) to magnitudes of the wave and tangential stresses are already not detailed.

It should be noted that the writing momentum balance equation in form (14) is not new and, as mentioned above, is widely used (Zaslavskii 1995; Makin and Kudryavtsev 1999; Polnikov 2009; among others). It only states the fact that contributions of the terms in the left-hand side of equation (12) into the left-hand side of (14) are not a priori distinguishable. Equations (12) and (14) are not ascribed to a certain vertical coordinate; they are valid throughout the wave-zone, as well as the wave spectrum is not tied to a specific level of the wavy interface.

Next, as usual, we assume that wave stress $\tau_w$ in (14) can be determined via the pumping function (4) by choosing an adequate representation for empirical wave-growth increment $\beta(\omega,\theta,W,\theta_w)$. But the formula for calculating the value of tangential stress $\tau_t$ is not known in advance. In this case, identification of function $\tau_t(u_*,S)$, apparently, must be performed by the methods of statistical fluid mechanics, i.e. by means of its parameterization via dimensional and dimensionless characteristics of the interface as a whole. This is the specificity of the new interpretation of equation (14) (see details below).

The value $\tau_{tot}$, standing in the right-hand side of (14), has a meaning of the total momentum flux from the wind to wavy surface, averaged over the wave ensemble. It is quite natural to assume that this momentum flux $\tau_{tot}$ has the value actually measured at some horizon in the WBL, located highly from the largest wave crests; that is, as usual,

$$\tau_{tot} = < -u^{'}w^{'} > \big|_{z>>H} = u_*^2 , \qquad (15)$$

where $u_*$ is the friction velocity.

It is also natural to assume that the both components of the total momentum flux, $\tau_w$ and $\tau_t$, depend on some principal characteristics of the system, such as: the wind speed at a fixed standard horizon (usually z = 10m), W (or its equivalent in the form of friction velocity $u_*$), the two-dimensional spectrum of waves $S(k_x,k_y)$ (or its equivalent in the frequency-angle representation $S(\omega,\theta)$). They could as well depend on such integral wave characteristics as: the peak frequency of the wave spectrum $\omega_p$, the mean wave height *H*, and a number of dimensionless characteristics of the system: the wave age $A = c_p / W$, and the mean wave slope $\varepsilon = Hk_p$; where $c_p$ and $k_p$ is the phase-velocity and wave-number of the wave component corresponding to the peak frequency, respectively. In this formulation, the problem of WBL-model realization is reduced to finding solution of equation (14) with respect to unknown value $u_*$ being a function of all the abovementioned parameters of the system.

## 3. Parameterization of $\tau_w$

Reasonability of extraction of wave stress $\tau_w$, associated with the energy transfer from the wind to waves, from total stress $\tau_{tot}$ is due to the fact that the value of $\tau_w$ can be measured to a certain extent, and even can be theoretically described in terms of the wave-dynamics equations within certain limits (see references in Phillips (1981); Komen et al. (1994), Donelan et al. (2006), Chalikov and Rainchik (2010)). It is this scrutiny of features for $\tau_w$ allows us to give its analytical representation in the form of (4), where the pumping function $IN$ is used, mathematical representation of which has widely accepted theoretical and empirical justification.

The most detailed representation of $IN$, useful for further, can be written in the form (Komen et al. 1994; Polnikov 2009)

$$IN(...) = \left(u_* / c_\omega\right)^2 (\rho_a / \rho_w) \tilde{\beta}(u_* / c_\omega, \theta, \theta_W) \cdot \omega S(\omega, \theta) \propto u_*^2 \propto W^2 , \qquad (16)$$

where the standard notation are used, which are not in need of clarification (see below). Physical content of $IN$ is the rate of the energy flux from the wind to waves (per unit surface area). The well-justified principal features of function *IN*(…) are as follows: 1) explicit quadratic dependence on friction velocity $u_*$ (or wind *W*), what especially extracted in writing the right-hand side of (16); 2) proportionality to the ratio of densities of air and water $(\rho_a / \rho_w)$; 3) linearity on wave spectrum *S*. The rest theoretical uncertainties of dimensionless function $\tilde{\beta}(u_*/c_\omega,\theta,\theta_W)$ are usually parameterized on the basis of various experimental measurements or numerical calculations, due to what its analytical representation is not uniquely defined (Komen et al. 1994; Donelan et al., 2006; Chalikov and Rainchik, 2010).[5]

In a narrow domain of frequencies $\omega$ (or wave numbers *k*), covering the energy-containing interval of gravity waves within the limits (the so called low-frequency domain)

$$0<\omega<5\omega_p, \tag{17}$$

function $\tilde{\beta}(u_*/c_\omega,\theta,\theta_W)$ with an accuracy of about 50% can be represented in the form (Yan 1987)

$$\tilde{\beta}(...)=32\{[1+0.136/(u_*/c_\omega)+0.00137/(u_*/c_\omega)^2]\cos(\theta-\theta_w)-0.00775(u_*/c_\omega)^2\}, \tag{18}$$

if the parameterization of $\tilde{\beta}(...)$ via $u_*$ is used. There are also present parameterizations of $\tilde{\beta}(...)$ via *W*, among which it is appropriate to mention one of the last (Donelan et al. 2006):

$$\tilde{\beta}(...)=G[\varepsilon(k),\theta,\theta_w](1-c_\omega(k)/W)^2, \tag{19}$$

where $G[\varepsilon(k),\theta,\theta_w]$ is the empirical function of the wave component steepness, $\varepsilon(k)$, the direction of its propagation, $\theta$, and the local wind direction, $\theta_w$.

In a higher and broader frequency domain, which includes the capillary waves ranging up to 100 rad/s (where the most significant part of the wave momentum flux is accumulated), the contact measurement of $\tilde{\beta}(u_*/c_\omega,\theta,\theta_W)$ becomes impossible due to evident technical reasons. This deficit of information can be compensated by another means, at certain extent. Namely, remote sensing and numerical estimates indicate the fact that, in the mentioned frequency domain, $\tilde{\beta}$ has a shape of a broad asymmetric "dome" with the maximum at frequencies of about 3rad/s (corresponding to wave numbers of the order of 1m$^{-1}$). Function $\tilde{\beta}$ drops to zero both at the low frequencies corresponding to condition $W/c_\omega$<1 (i.e. when the waves overtake the wind) and at high frequencies corresponding to the region of the phase-velocity minimum (Elfouhaily et al. 1997; Kudryavtsev et al. 2003) (Fig. 2).

[5] There are more detailed representation of $\tilde{\beta}$, using dependences on wave age *A* and wave component steepness $\varepsilon(k)$ ( Donelan et al., 2006). But they are not discussed here.

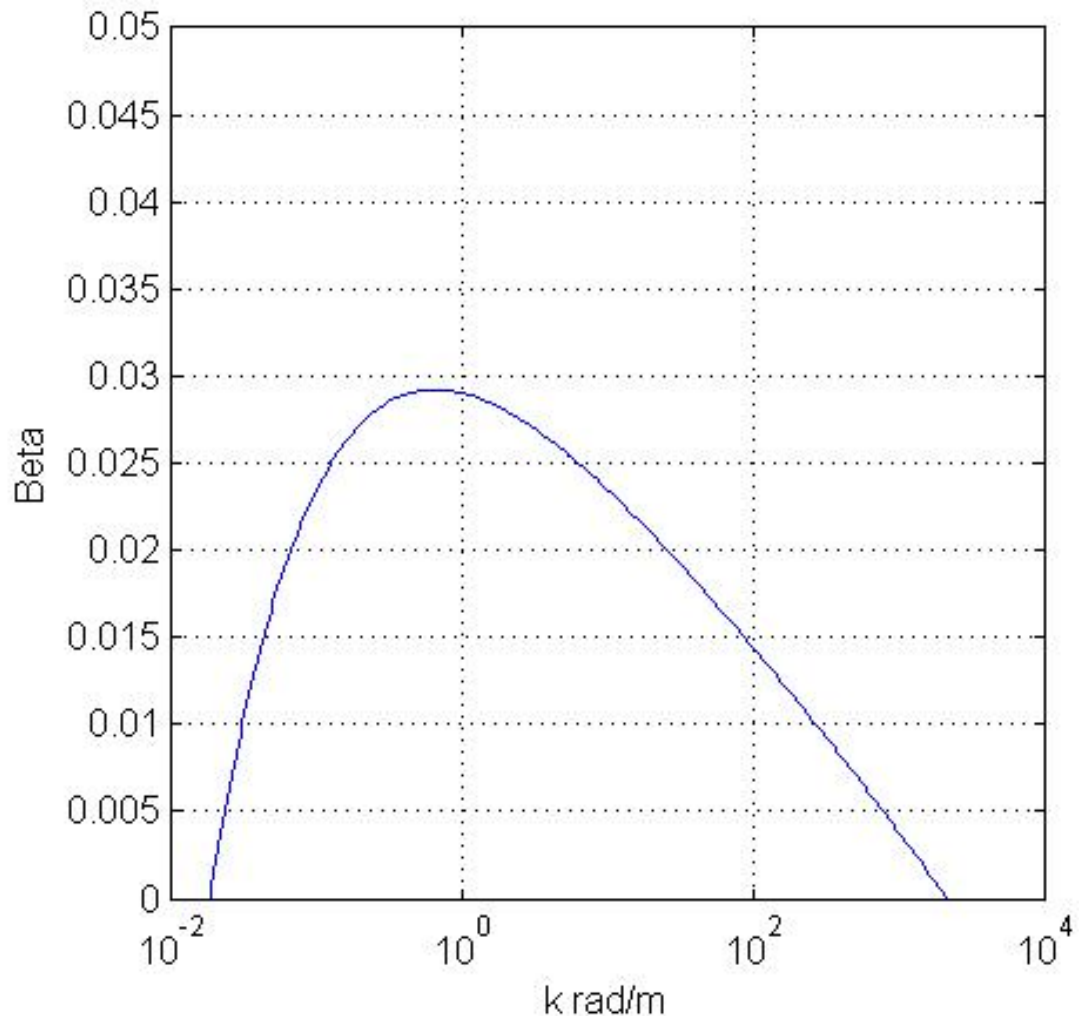

Fig. 2. The general form of the normalized growth increment $\tilde{\beta}(k)$ in the wide band of wave numbers.

Angular dependence of function *IN (...)* is less studied. Usually, it is parameterized by the cosine of the angle difference $(\theta - \theta_W)$, where $\theta_W$ is the direction of the local wind (see details in references). Nevertheless, these data are sufficient to solve our problem.

Additionally, regarding to estimates of $\tau_w$ based on formula (4), it should be noted that the integral in the right-hand side of (4) is slowly convergent (Polnikov 2009; Zaslavskii 1995; Makin and Kudryavtsev 1999; among others). It is for this reason the limits of integration in (4) should cover the domain from the maximum frequency of the wave spectrum $\omega_p$, having the order of (0.5-1)rad/s, up to frequencies of about 100 rad/s ($k \cong$ 1000rad/m), including the domain of capillary waves existence where $\tilde{\beta}$ is close to zero. The using this extended domain ensures convergence of the integral in (4). However, the wave spectrum in such domain is very difficult to obtain both by contact measurements and by regular numerical calculations. For this reason, a sufficiently accurate assessment of flux $\tau_w$ requires using a special approach. It consists in sharing the wave spectrum into two fundamentally different parts: the low-frequency spectrum $S_L$ of gravity waves (LFS), actually measured (or calculated by some numerical model), and the high-frequency spectrum $S_H$ (HFS), a modeling representation of which can be found by using a variety of remote measurements and their theoretical interpretation (Elfouhaily et al. 1997; Kudryavtsev et al. 2003). Thus, the farther progress in constructing the WBL-model consists in using representation of the wave spectrum in the form of two terms, for example, in the form proposed in (Elfouhaily et al. 1997 ):

$$S(\omega,\theta) = S_L \cdot cut + S_H \cdot (1 - cut), \qquad (20)$$

where $cut(A,\omega)$ is the known cutoff factor.

Since we are interested in integrated estimations of the type of ratio (4), in the proposed WBL-model it is quite acceptable to represent $cut(A,\omega)$ in the form of a step function changing abruptly from 1 to 0 at the point $\omega \approx 3\omega_p$ while the frequency increasing. Accepting that the wave phase-velocity is given by the general ratio: $c_\omega = \omega(k)/k = [gk(1+\gamma k^2)\mathrm{th}[kd]]^{1/2}/k$, and the wind direction is $\theta_W = 0$, formula (4) can be rewritten in the most general kind as

$$\tau_w = \rho_a u_*^2 [\int_{\Omega_L} \frac{k^2}{\mathrm{th}[kd]} \cos(\theta)\tilde{\beta}(...)S_L(\omega,\theta)d\omega d\theta + \int_{\Omega_H} \frac{k^2\cos(\theta)}{(1+\gamma k^2)\mathrm{th}[kd]}\tilde{\beta}(...)S_H(\omega,\theta)d\omega d\theta] = \\ = \tau_{wL} + \tau_{wH}, \qquad (21)$$

where $\gamma$ is the water surface tension normalized on the gravity acceleration *g*, and *d* is the local depth of the water layer.

Thus, the explicit shearing the wave-part stress into two summands is introduced in (21): $\tau_{wL}$ is the LFS-contribution provided by the relatively easily measured energy-containing part of the wave spectrum, estimated in the integration domain $\Omega_L$, and $\tau_{wH}$ is the HFS-contribution accumulated in the integration domain $\Omega_H$. Herewith, for HFS it is necessary to accept a modeling representation based on generalization of a large number of remote-sensing observations of various kinds (Elfouhaily et al. 1997; Kudryavtsev et al. 2003). Assuming that the both parts of the total (gravity-capillary) wave spectrum can principally be represented in a quantitative form, one can state that the magnitude of the wave-part stress is known as a function the following arguments: wave spectrum *S*, wind speed *W*, friction velocity $u_*$, wind direction $\theta_W$, peak-wave-component direction $\theta_p$, wave age $A$, and mean wave-field steepness $\varepsilon$. Introducing the dimensionless variables, $\hat{\tau}_{wL}$ and $\hat{\tau}_{wH}$, we can write the ratio

$$\tau_w = u_*^2[\hat{\tau}_{wL}(W, u_*, S, A, \varepsilon, \theta_p, \theta_w) + \hat{\tau}_{wH}(u_*, S, A, \varepsilon, \theta_p, \theta_w)], \qquad (22)$$

the summands of which, $\hat{\tau}_{wL}$ and $\hat{\tau}_{wH}$, are potentially known. Herewith, we especially reserve the wind speed *W* as the argument, to keep the possibility of getting dependence $u_*$ on $W$ as the solution of equation (14) with respect to unknown function $u_*(W,S)$.

Besides the said, to simplify the procedure of assessing the high-frequency component $\hat{\tau}_{wH}$, it is proposed to tabulate the numerical representation of the latter, calculated on the basis of known shape-function for the HFS (Elfouhaily et al. 1997; Kudryavtsev et al. 2003).

## 4. Numerical Specification of $\hat{\tau}_{wH}$.

With the aim to specify function $\hat{\tau}_{wH}(u_*, A, ...)$, let us calculated it numerically as a function of friction velocity $u_*$ and wave age *A*, defining *A* via the wave number of dominant waves, $k_p$. For this purpose, we use the modeling representation of HFS spectrum $S_H(k,\theta,\theta_w,u_*)$ from (Kudryavtsev et al., 2003) where the matter is disclosed with a maximum specificity. Since, according to the reference, the analytical presentation of $S_H(k,\theta,\theta_w,u_*)$ is extremely cumbersome, to save the space, here we confine ourselves to graphic illustrations of $\hat{\tau}_{wH}$, made by the method described in (Kudryavtsev et al., 2003). This approach allows us to obtain the quantitative representation of function $S_H(k,\theta,\theta_w,u_*)$ and the sought function $\hat{\tau}_{wH}(u_*, A)$ [6].

---

[6] In Fig. 3 the full calculated spectrum $S(k,u_*)$, given by formula (20), contains the LFS-part calculated following to its parameterization in Donelan et al (1985).

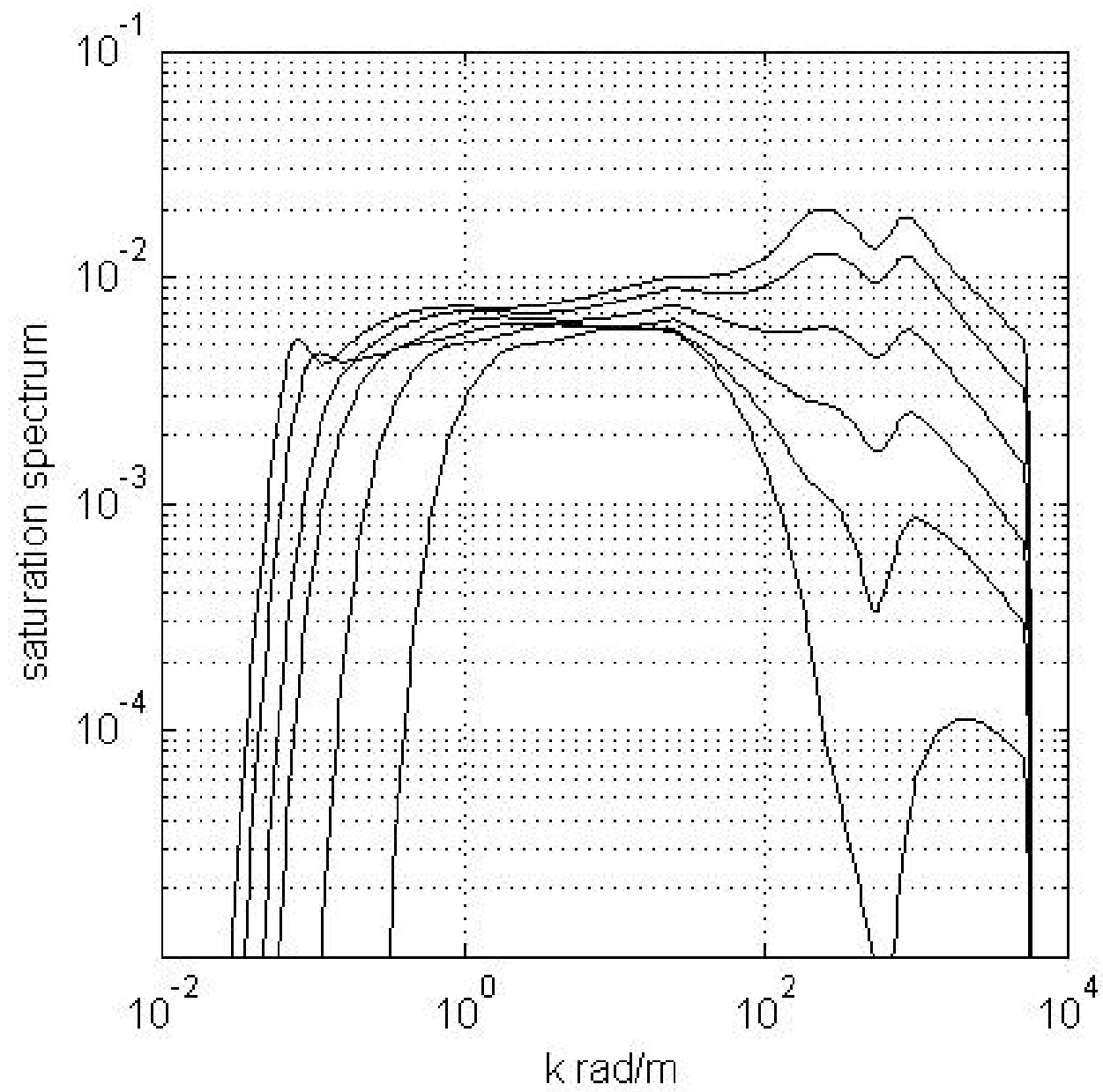

Fig. 3. Saturation spectrum $B(k,u_*)$ in the wide band of wave numbers *k*.
Growth of the level for a high-frequency part of $B(k,u_*)$
reflects its dependence on friction velocity $u_*$ changing from 0.1 up to 1m/s.

So, in Fig. 3 the modeling one-dimensional saturation spectrum (or the curvature spectrum) $B(k,u_*) = k^3 S(k,u_*)$ is presented for a set of values of friction velocity $u_*$ varying from 0.1 to 1 m/s. The spectrum is calculated in the wave-numbers band $0.01 < k < 10^4$ rad/m ; the part of the shown spectrum, in the band of wave numbers *k*<3rad/m, belongs to the LFS given by formulas in Donelan et al. (1985). From Fig. 3 it is clearly seen that in the HF-band of wave numbers *k*, occupying the range from 100 to 1000rad/m, the HFS depends very strongly on friction velocity $u_*$. However, in the intermediate range (10-50 rad/m), the intensity of $B(k,u_*)$ varies much weaker. This, numerically established fact indicates the weak link of the HFS-intensity $S_H(k,\theta,u_*)$, with the LFS-intensity $S_L(k,\theta,u_*)$, allocated in the domain $k \le 3$rad/m. It is this fact that allows us to evaluate independently fluxes $\hat{\tau}_{wL}$ and $\hat{\tau}_{wH}$, accepting representation (21) for the total stress $\hat{\tau}_w$.

The numerical calculations of the flux function $\hat{\tau}_{wH}(u_*,k_p)$, performed with the use of representations of HFS $S_H(k,\theta,u_*)$ and increment $\tilde{\beta}\left(u_*/c_\omega,\theta,\theta_w\right)$ taken from (Kudryavtsev et al., 2003), are shown in Fig. 4. The lines correspond to different (fixed) values of the peak wave number $k_p$: the upper line corresponds to values of $k_p$ in the range $0.1 \le k_p \le 1$rad/m, while the low-lying lines correspond to a set values of $k_p$ increasing with step of 0.2rad/m from 1 to 4 rad/m (bottom line). Calculations were performed by varying the value of $u_*$ in the range from 0.06 to 1.05m/s, actually covering the entire admissible range of friction velocity values.

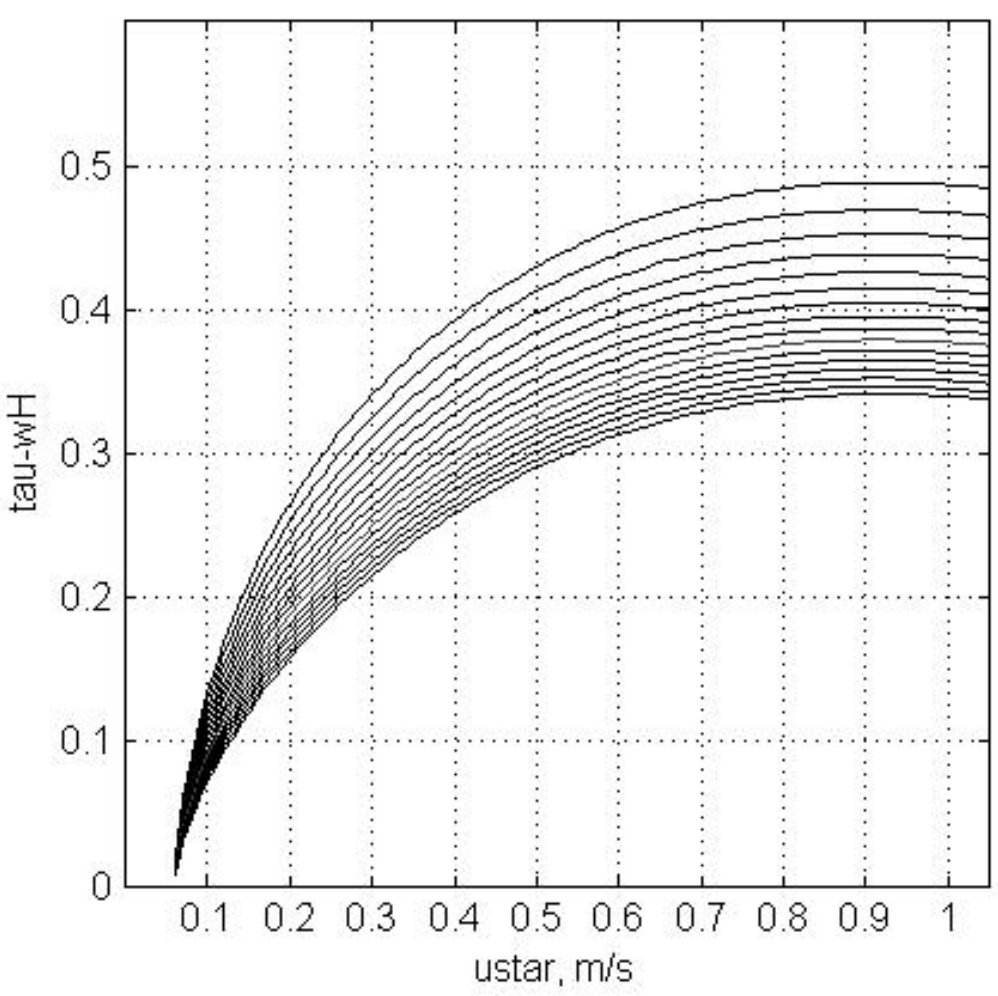

Fig. 4. Estimation of function $\hat{\tau}_{wH}(u_*, k_p)$, obtained for the modeling HFS taken from( Kudryavtsev et al. 2003).

From Fig. 4 it follows that flux $\hat{\tau}_{wH}(u_*, k_p)$ grows rapidly with $u_*$ for all values of $k_p$ and reaches a plateau for values $u_* > 0.8$m/s, indicating the limiting contribution of HFS to the full normalized stress $\hat{\tau}_w$. As seen, for $k_p > 1$rad/m the level of this contribution depends on the value of peak wave number $k_p$: the higher $k_p$ the lower a percentage of the HFS-contribution. A detailed examination showed that this effect is due to the above-described behavior of the growth increment $\tilde{\beta}$ (see Figure 2). Indeed, while increasing $k_p$, the lower limit of integration domain $\Omega_H$ in (21) moves up in the wave number scale, and the corresponding value of $\tilde{\beta}$ begins to decline, providing less weighting factor under the second integral in (21), what leads to the effect under consideration.

The observed numerical effect does somewhat complicate the problem of solving equation (14), because the magnitude of $\hat{\tau}_{wH}$, as it would seem, is to be determined mainly by the shape of the HFS, i.e. depends on the friction velocity $u_*$, only. However, Fig. 4 makes it clear that there is also a dependence of $\hat{\tau}_{wH}$ on $k_p$, which is naturally determined by the choice of cutoff factor $cut(A,\omega)$ in (20). Consequently, the choice of the form of $cut(A,\omega)$ is one of the fitting elements of proposed WBL-model. However, such a choice is not principal in the physical sense, rather it is only a technical element of the model. Therefore, it is quite acceptable the choice of function $cut(A,\omega)$, providing an abrupt transition of the LFS to the HFS at wave numbers of the order of 1rad/m.

The analysis performed suggests the following specification of function $\hat{\tau}_{wH}(u_*,k_p)$. In view of the smoothness of function $S_H(k,\theta,u_*)$ and practical independence of its shape of the form for $S_L(k,\theta,u_*)$, in the practical version of WBL-model it is not necessary each time to recalculate the amount of $\hat{\tau}_{wH}(u_*,k_p)$, making loops over a set of fixed values of $u_*$ and $k_p$. It is enough to calculate this function once, in terms of the above-described scheme, for a practically important set of values $u_*$ and $k_p$, and simply to tabulate the values obtained (for example, those which are shown in Fig. 4). By this way we close the issue of parameterization of $\hat{\tau}_{wH}(u_*,k_p)$ completely. It is this approach will be realized below in Section 6 of the paper.

Now, to finalize equation (14) specification, it requires only to specify the form of function $\tau_t(u_*,A,c_p,...)$.

## 5. Parameterization of $\hat{\tau}_t$.

In contrast to flux $\tau_w$, the value of tangential flux $\tau_t$, transferring the momentum to the underlying surface as a whole, is practically not determined in the light of equation (12) derivation. To this regard, it is appropriate to note that in ocean-circulation models, ignoring the existence of waves at the interface, the value of $\tau_t$ is associated with the full wind stress $\tau_{tot}$, as the source of drift currents. In fact, besides from the drift currents, the wind generates waves. Therefore, there is a large and physically justified difference between tangential stress $\tau_t$ and full stress $\tau_{tot}$.

Due to features of equation (12) derivation and its representation in the form of (14), the theoretical specification of expression for $\tau_t$ is principally difficult in the case of real stochastic wavy interface. In particular, it is due to the lack of mathematical algorithms for implementation of the above-mentioned averaging over the wave-field ensemble. Experimental determination of function $\tau_t(u_*, A, \varepsilon, S)$, apparently, has a prospective, if one takes into account possibility of measuring $u_*$ and using the calculated values of $\tau_w(u_*, A, \varepsilon, S)$, obtained for synchronous measurements of $S(\omega, \theta)$. However, such estimates of function $\tau_t(u_*, A, \varepsilon, S)$ are not known for us. Therefore, in view of uncertainty of function $\tau_t(u_*, A, \varepsilon, S)$ representation, in our version of WBL-model this function will be built phenomenologically on the basis of the special analysis of the wind profile in the wave-zone, performed earlier in (Polnikov 2010).

First of all, note that the accounting both the air kinematic viscosity *v* and the standard log-profile for mean wind $W(\mathrm{z}) \propto u_* \log(z / z_0)$ is not physically correct in constructing the parameterization of $\tau_t(u_*, A, \varepsilon, S)$, as the latter should be valid in the space of wave-zone distributed in height from -*H* to *H*, counted with respect to the mean surface level. The former attributes are applicable only in the case of quasi-stationary in time, and horizontally quasi-homogeneous interface. In the wave-zone, all these attributes of "the hard-wall turbulence" disappear in the course of ensemble-averaging the integrated balance equation (12) (see Section 2). Therefore, any additional information about the wind profile in the wave-zone is needed. The hint in constructing parameterization of $\tau_t(u_*, A, \varepsilon, S)$ is provided by the following numerical fact (Polnikov 2010).

The author has recently processed the data of numerical experiments by Chalikov and Rainchik (2010), devoted to the joint simulation of the wind and wave fields in the case of air flow over the wavy surface. Kindly provided data of the simulated wind field *W(x,z)*, obtained by the mentioned authors in the curvilinear coordinates following to the wavy interface, have been converted to the Cartesian coordinates, and the profile of mean wind *W*(z) was built in the wave-zone, i.e. in the area between troughs and crests of waves where water and air present alternately. Details of these calculations can be found online at the site ArXiv.org (Polnikov 2010). It turned out that in the wave-zone, the mean wind profile *W(z)* has a linear dependence on z (Fig. 5). Herewith, linear profile *W(z)* is spread from -*D* to 3*D* in the height measured relative to the mean water level assumed to be zero. Here, *D* is the standard deviation of the wavy surface, given by (13) [7].

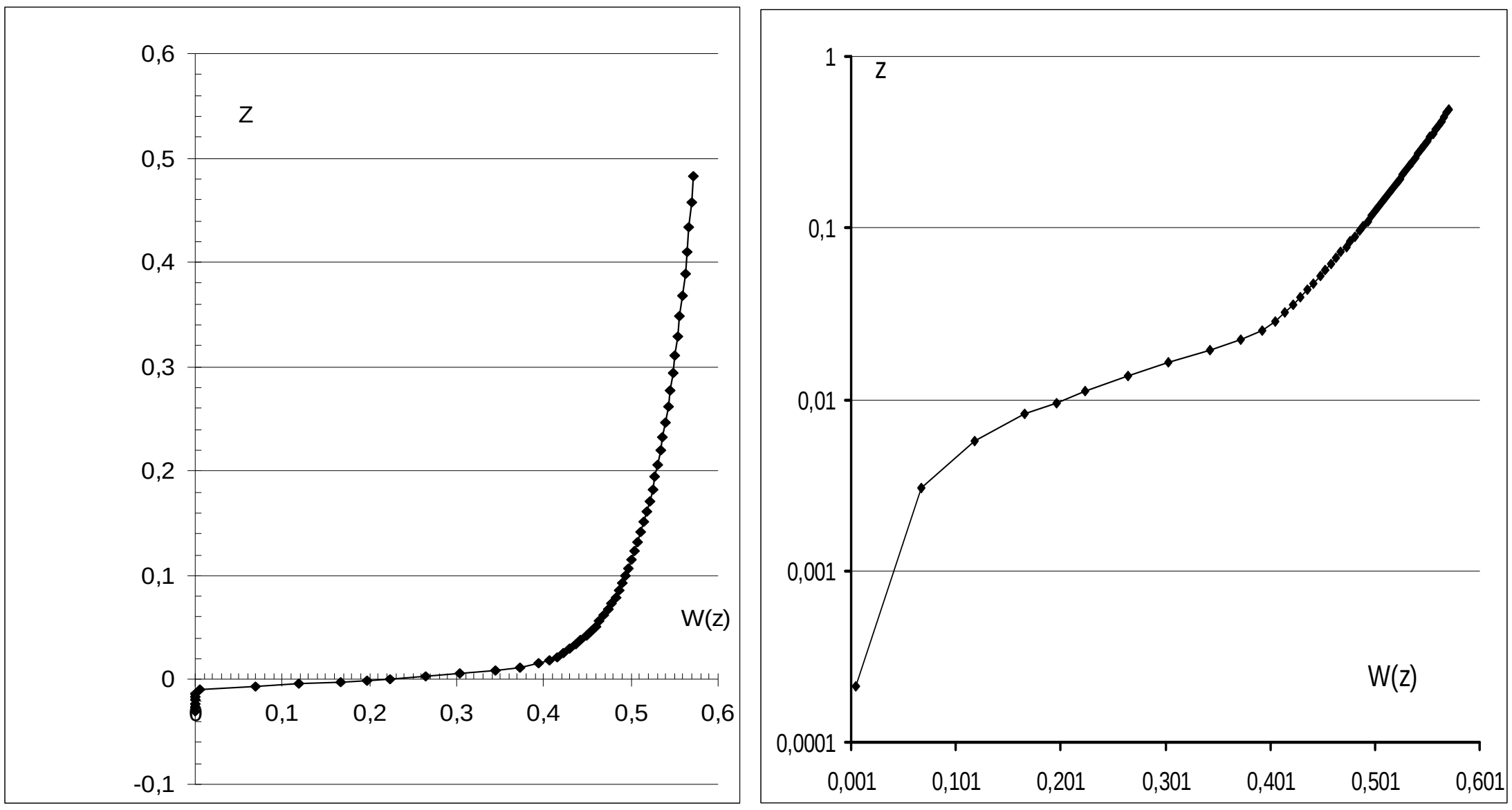

a) b)

Fig. 5. Calculated mean-wind profile *W*(*z*) over wavy surface (following to Polnikov 2010):
a) linear coordinates; b) semi-logarithmic coordinates .
Results are shown in the dimensionless units; zero value of *z* corresponds to the mean surface level $\langle \eta \rangle$ .

---

[7] Though, starting from height 3*D* over the mean surface level $\langle \eta \rangle$, the mean wind profile *W(z)* becomes close to the logarithmic one, in full accordance with the experimental observations (Drennan et al. 1999) (see Fig 5b). Herewith, upper the wave-zone, the wave-part flux $\tau_w$ has the exponential decay law close to the traditional representation as $\tau_w(z) \propto \exp(-k_p z)$ (Polnikov, 2010).

From the results by Polnikov (2010), in particular, it follows that in the wave-zone ([$-D < z < 3D$]) the vertical gradient of the mean wind profile is given by

$$\frac{\partial W(z)}{\partial z} \cong \frac{c_p}{H}, \tag{23}$$

where $c_p$ is the phase velocity of dominant waves, $H$ is the average wave height.

Basing on the said, and in light of the fact that the linear wind profile is precisely corresponding to the viscous-flow sublayer (Monin and Yaglom 1971), one may assume that in the case considered, the wave-zone plays a role of the viscous layer of the interface (replacing the traditional viscous sublayer defined by the kinematic viscosity of air: $d_\nu \approx 10\nu / u_* \ll D$). Thus, considering the wave-zone as a kind of generalization of the viscous layer in the case of the wavy interface, and attracting in this zone the well-known theoretical expression (5) for turbulent stress, the analytical representation for the sought function $\tau_t(u_*, A, \varepsilon, S)$ can be obtained by writing explicit expressions both for velocity gradient $\partial W(z) / \partial z$ and for a constant value of turbulent viscosity $K$, which are realized in the wave-zone. This is the main new and fundamental feature of the proposed WBL-model.

Estimation of $\partial W(z) / \partial z$ follows from (23), and the closure of parameterization for $\tau_t(u_*, A, \varepsilon, S)$ requires only a formula for turbulent viscosity $K$, which is valid in the wave-zone. Herewith, for estimation of $K$ it is quite acceptable to consider it as the value $K_w$ independent of the vertical coordinate $z$, which is realized at given wind-wave conditions. Since, in the wave-zone, the characteristic scale of the vertical velocity of air is the friction velocity $u_*$, and the space-scale of vertical motion is determined by the wave height $H$, from dimensional considerations it follows

$$K = K_w = (u_* H) \cdot fun(\varepsilon, A, ...) \quad , \tag{24}$$

where $fun(\varepsilon, A, ...)$ is the unknown dimensionless function of the integrated wave parameters, determined in the course of the fitting process during the model verification.

In the frame of the proposed approach, we believe that the wave-age parameter $A$, being associated only with the ratio of the dominant waves phase-velocity to the wind speed, is less crucial for evaluation of $K_w$ than the mean steepness of waves $\varepsilon$. Therefore, at present, preliminary stage of the WBL-model specification, the account of dependence $K_w(A)$ may be excluded, and the sought function can be written in the form: $fun(\varepsilon) \propto \varepsilon^n$, where $n$ is the fitting integral power. By this way we are accepting minimal assumptions.

As a result, the set of formulas (5), (14), (21) - (24) completes the theoretical basis of the proposed WBL-model construction, and we can proceed with its specification based on the verification procedure carried out by means of comparison model calculations with observations.

## 6. Method and Results of the Model Verification

### 6.1. Verification Method

The procedure of verification consists in checking correspondence of empirical values of friction velocity $u_*$ to the calculated values $u_{*mo}$ following from the solution of equation (14). Herewith, the values of friction velocity $u_*$ and spectrum $S(k,\theta)$ should be measured simultaneously. According to the literature, such kind of measurements are rather common (see references in Makin and Kudryavtsev 1999; Kudryavtsev and Makin 2001; Babanin and Makin 2008). Nevertheless, they are hardly available for us, and this is the main technical difficulty of performing the verification procedure. Results of simultaneous measurements $S(\omega,\theta)$ and $u_*$, being at our disposal, are very limited and based completely on the data described in (Babanin and Makin 2008), kindly provided by Babanin.

After preliminary selection of the most representative data, in this paper we have attracted to calculations only 26 of 72 series of measurements made in 2002-2004 in shallow water: Lake George in Australia. For this reason, the dispersion relation for waves in the finite depth water was used (for legend, see formula 21):

$$\omega^2(k) = gk(1+\gamma k^2)\mathrm{th}(kd) \ . \qquad (25)$$

Full information about the measured parameters of the "wave-boundary layer" system is described in the reference; so, it is further provided in an extremely compressed form (Table 1) sufficient to analysis of verification results.

6.2. Specification of the Model

The model is given by equation (14) which is easily solved with respect to $u_*$ by the standard method of dividing the interval in half. In this case, the low-frequency part of the wave momentum flux $\tau_{wL}(W, u_*, S_L)$ is calculated on the basis of the LFS measurements: $S_{ex}(\omega,\theta)$ (i.e. $S_L = S_{ex}(\omega,\theta)$ ), by the formula

$$\hat{\tau}_{wL} = u_*^2 \cdot \int_{\Omega_L} \frac{k(\omega)^2 \cos(\theta)}{\mathrm{th}[k(\omega)d]} \tilde{\beta}(...) S_L(\omega,\theta) d\omega d\theta \ . \qquad (26)$$

The denominator under the integral in (26) means accounting a finiteness of the water layer depth *d*; domain of integration $\Omega_L$ is limited in frequency band by the value $\omega_{\max} = \omega(k_g)$ corresponding to $k_g$ = 1rad/m; and the growth increment is accepted in the representation[8]

$$\tilde{\beta}(...) = 40 u_{*0}^2 \{[1+0.136/(u_{*0}/c_\omega)+0.00137/(u_{*0}/c_\omega)^2]\cos(\theta-\theta_p) - 0.00775(u_{*0}/c_\omega)^2\} , \qquad (27)$$

where we postulate that

$$u_{*0}^2 = C_{d0} W^2 \qquad (28)$$

with the fitted value $C_{d0}$ equal to 0.002. The phase velocity $c_\omega = \omega(k)/k$ is following from (25).

The choice of $\tilde{\beta}(...)$ in form (27) - (28) is caused only by technical reasons (see footnote 8), i.e. other options are possible. In our model it is only important that $\tilde{\beta}(...)$ is expressed via the wind at the standard horizon *W*. In physical terms, this choice constitutes acceptance of the postulate that the transfer of momentum from the atmosphere to the gravity waves is determined by the value of mean local wind *W*. From a technical standpoint, this postulate is needed to ensure that in the balance equation (14), written for friction velocity $u_*$, wind *W* is also present. This approach provides the sought dependence $u_*(W)$. From physical point of view, this approach is fully justified, since wind *W* is the primary source of both the wave energy and the statistical structure of the interface.[9] In this version, of course, $C_{d0}$ is the parameter of the model, the value of which is determined during verification.

---

[8] Expression in figure brackets of (27) corresponds to the generalized parameterization of empirical $\tilde{\beta}(...)$, given by formula (18) and proposed by Yan(1987). This form is widely used by the author (see references in Polnikov 2009). In the present case, approximation (27-28) is tentative but not the final one.

High-frequency part of the wave flux $\tau_{wH}(u_*,k_p) = u_*^2 \cdot \hat{\tau}_{wH}(u_*,k_p)$ is calculated along the line described in Section 4. Here, as usual, it is believed that the momentum flux to the high-frequency waves is mainly defined by the friction velocity rather then the wind at the standard horizon (Elfouhaily et al. 1997; Kudryavtsev et al. 2003). This replacement of wind *W* to friction velocity $u_*$ is caused by the much smaller scales of the high-frequency waves and the proximity of the matching layer to the interface. Herewith, we make a pre-tabbing the two-dimensional array $\hat{\tau}_{wH}(u_*,k_p)$, choosing $cut(k_g,k)$ as a step-function changing abruptly from 1 to 0 at the frontier $k \ge k_g = 1$. The analogous representation of two-dimensional matrix $\hat{\tau}_{wH}(u_*,k_p)$ is shown in Fig. 4 (as far as the table-representation is too cumbersome).

And finally, according to the ideology of Section 5, the tangential stress is given as

$$\tau_t = (u_* c_p) \cdot c_t(n) \varepsilon^n . \qquad (29)$$

In this version of the model, the most proper values: $n$ = 1 and $c_t(1)$=0.8, are found in the verification process.

---

[9] Note that with the account of the said in Section 3 about uncertainty of function $\tilde{\beta}(...)$, the postulate accepted does not introduce any new assumptions (i.e. additional to the known) about the physics of the interface dynamics. Moreover, parameterizations of $\tilde{\beta}(...)$ via *W* are widely spread (Donelan et al., 2006).

Table 1. Specification of the measurements data and the results of the WBL-model verification.

| № of run | № of data series | $d$, m | $W_{10}$, m/s | $u_*$, m/s | $f_p$, Hz | $k_p$, 1/m | $10\varepsilon = 10k_pH$ | $A = c_p/W$ | $10^3*Cd_E$, experiment | $10^3*Cd_M$, model | Drag coeff. deficit: $[Cd_M/Cd_E-1]$, % |
|---|---|---|---|---|---|---|---|---|---|---|---|
| 1 | 010004 | 0.86 | 9.9 | 0.38 | 0.51 | 1.3 | 0.91 | 0.25 | 1.47 | 1.55 | +06 |
| 2 | 010055 | 0.87 | 11.8 | 0.51 | 0.41 | 1.0 | 0.84 | 0.22 | 1.87 | 1.45 | -22 |
| 3 | 010248 | 0.93 | 14.8 | 0.67 | 0.37 | 0.8 | 1.00 | 0.19 | 2.05 | 1.81 | -12 |
| 4 | 111051 | 1.14 | 12.9 | 0.57 | 0.61 | 1.6 | 1.33 | 0.19 | 1.95 | 2.24 | +15 |
| 5 | 111156 | 1.14 | 12.6 | 0.57 | 0.58 | 1.4 | 1.28 | 0.20 | 1.95 | 2.19 | +12 |
| 6 | 111224 | 1.14 | 11.9 | 0.50 | 0.61 | 1.6 | 1.27 | 0.20 | 1.77 | 1.84 | +04 |
| 7 | 111402 | 1.14 | 13.0 | 0.56 | 0.58 | 1.4 | 1.26 | 0.19 | 1.86 | 1.99 | +07 |
| 8 | 111538 | 1.14 | 11.6 | 0.48 | 0.63 | 1.6 | 1.31 | 0.20 | 1.71 | 2.00 | +17 |
| 9 | 141215 | 1.09 | 10.1 | 0.39 | 0.73 | 2.2 | 1.17 | 0.21 | 1.49 | 1.49 | 00 |
| 10 | 141250 | 1.09 | 11.0 | 0.45 | 0.71 | 2.0 | 1.25 | 0.20 | 1.67 | 1.75 | +05 |
| 11 | 141305 | 1.00 | 14.1 | 0.64 | 0.37 | 0.8 | 0.88 | 0.20 | 2.06 | 1.41 | -31 |
| 12 | 151249 | 1.10 | 9.9 | 0.36 | 0.37 | 0.8 | 0.78 | 0.30 | 1.32 | 1.40 | +06 |
| 13 | 151318 | 1.10 | 9.40 | 0.35 | 0.37 | 0.8 | 0.64 | 0.32 | 1.39 | 0.89 | -36 |
| 14 | 151342 | 1.10 | 9.10 | 0.33 | 0.44 | 1.0 | 0.87 | 0.31 | 1.32 | 1.83 | +39 |
| 15 | 151410 | 1.1 | 9.7 | 0.33 | 0.41 | 1.0 | 0.88 | 0.29 | 1.16 | 1.78 | +54 |
| 16 | 161425 | 0.71 | 6.9 | 0.26 | 0.93 | 3.5 | 1.02 | 0.24 | 1.42 | 1.31 | -07 |
| 17 | 201446 | 0.89 | 6.7 | 0.26 | 1.02 | 4.2 | 1.36 | 0.23 | 1.51 | 1.51 | -00 |
| 18 | 261148 | 0.84 | 5.9 | 0.28 | 1.02 | 4.2 | 1.11 | 0.26 | 2.25 | 1.01 | -54 |
| 29 | 271100 | 0.95 | 6.1 | 0.28 | 0.49 | 1.2 | 0.68(!) | 0.42 | 2.11 | 1.18 | -44 |
| 20 | 281544 | 0.96 | 5.1 | 0.16 | 1.02 | 4.2 | 1.11 | 0.30 | 0.98 | 1.11 | +13 |
| 21 | 311638 | 0.94 | 9.3 | 0.44 | 0.49 | 1.2 | 0.69(!) | 0.28 | 2.24 | 1.04 | -53 |
| 22 | 311823 | 1.12 | 19.8 | 0.98 | 0.34 | 0.8 | 1.04 | 0.15 | 2.45 | 2.24 | -08 |
| 23 | 311845 | 1.04 | 15.0 | 0.63 | 0.29 | 0.6 | 0.89 | 0.2 | 1.76 | 1.34 | -24 |
| 24 | 311908 | 0.93 | 12.9 | 0.53 | 0.32 | 0.70 | 0.91 | 0.22 | 1.69 | 1.56 | -07 |
| 25 | 312021 | 1.02 | 13.7 | 0.57 | 0.39 | 0.87 | 1.03 | 0.21 | 1.73 | 1.98 | +14 |
| 26 | 312111 | 0.86 | 9.3 | 0.35 | 0.41 | 1.0 | 0.83 | 0.28 | 1.42 | 1.34 | -06 |

Note. The column for wave steepness $\varepsilon$ is shaded for better identification.
Sign (!) in this column does mean “a mismatch” between values of $\varepsilon$ and $k_p$.

### 6.3. Results of Verification

Results of calculations for the set of 26 series of measurements are presented in Tab. 1. From the table it follows that the proposed version of WBL-model shows fairly good correspondence with measurements. Thus, according to the last column of Tab. 1, only three cases of 26 have the relative error $\rho_{Cd}$, defined by (8), exceeding the technically important quantity of 50%, being beyond the acceptable one in practical terms. The mean value of error $\rho_{Cd}$ is of the order of 15-20%, what is close to the accuracy of the measurements. It indicates a high representativeness of the model under consideration.

Nevertheless, it should be noted that the results reported here are not final yet, and still require a large series of comparisons based on a more extensive database of measurements, before the optimal choice of the used fitting functions and parameters of the model will be finally determined. Such work is planned in the nearest future.

In addition to the said, and in order to analyze the quality of experimental results presented in Tab. 1, one should pay attention to the following. Even in our selected series, sometimes there is a clear discrepancy in values between too small wave steepness $\varepsilon$ and very high values of the peak wave number $k_\mathrm{p}$ (see note to Tab. 1). It is in these cases, there are the most significant errors of the discussed model. The foregoing demonstrates the need of careful controlling the measurements accuracy and selection of empirical data involved to verification of such models.

## 7. Conclusion

Summarizing, we formulate the main regulations of the proposed approach to solving the problem of relation between friction velocity $u_*$, as the main parameter of the atmospheric layer, and the main characteristics of the wind-wave system: the local wind at the standard horizon W, and the two-dimensional spectrum of waves $S(\omega,\theta)$.

First. The derivation of the initial balance equations for the momentum flux at the wavy interface (Section 2) shows the need of rethinking interpretation of the components of balance equation (1), as far as the latter is to be averaged in the wave-zone covering the area between through and crests of waves (Fig. 1b). To solve this issue, it seems to be required not so experimental efforts but rather the detailed theoretical analysis of features of the wind-flow dynamics near a wavy surface, small-scale details of which can be obtained mainly by numerical simulation (for example, by analogy with simulations in (Li et al. 2000; Sullivan et al. 2000; Chalikov and Rainchik 2010; among others).

Second. It is proposed to share the total momentum flux from wind to wavy surface $\tau_{tot}$ into two principally different components, only: the wave part $\tau_w$ responsible for the energy transfer to waves, and the tangential part $\tau_t$ that does not provide such transfer.

Third. The point of calculating the wave part of the momentum flux $\tau_w$ can be regarded as the practically solved one, basing on the following known facts: (a) the function of energy transfer rate from wind to waves $IN(wind,S)$ is known and has the kind (16); (b) dependence of the HFS-shape on a wave age and friction velocity is given in (Kudryavtsev et al. 2003); and (c) the intensity of the low-frequency spectrum of waves (in the range 0 < k < $3 \div 4$ rad/m) can be easily estimated. Herewith, the contributions of LFS and HFS to the wave-part of momentum flux can be calculated independently.

Fourth. In order to relate friction velocity $u_*$ to wind speed at the standard horizon W in the balance equation, the following approach is accepted. In the course of calculating wave-part flux $\tau_w$, function $IN(wind,S)$ should be expressed via wind $W$ to calculate the LFS-contribution $\tau_{wL}$ (for example, as in formulas (26)-(28)), though while assessing contribution of the HFS, $IN(wind,S)$ is to be expressed via $u_*$. In view of the known uncertainties in description of the wave-growth increment function $\tilde{\beta}(u_* / c_\omega, \theta, \theta_W)$ (Komen et al. 1994; Donelan et al. 2006), the above-said regulation does not introduce any significant changes in the traditional understanding of the interface dynamics. It is simply a convenient technical tool allowing achieving the task solution.

Fifth. Regarding to tangential stress $\tau_t$, there is not commonly used and theoretically justified analytical representation for it in terms of the system parameters. Therefore, to describe $\tau_t$, in the present model it is proposed using the similarity theory. The main idea is based on the proposition that the wave-zone, located between troughs and crests of waves, is an analogue of the traditional friction layer. This idea is supported by the results of the author's analysis (Polnikov 2010) of the numerical simulations performed in (Chalikov and Rainchik 2010). According to this results, the mean wind profile $W(z)$ is linear in vertical coordinate $z$ what is typical for a vicious layer. Thus, by using formula (5), function $\tau_t$ can be parameterized simply via the mean-wind gradient $\partial W(z)/\partial z$ and a constant turbulent viscosity $K$, as functions of the system parameters (formulas (5), (21)-( 24)). As a result, the balance equation (14) becomes closed what provides the problem solution.

Sixth. The results of verification of the proposed model, performed on the database obtained in the shallow water (Babanin and Makin 2008), indicate a high representativeness of the constructed integrated WBL-model (Tab. 1). The mean value of the relative error for the drag coefficient, defined by (8), is about 15-20%, what is a fairly good result, taking into account the accuracy of measurements.

Thus, despite the limited empirical database, the results of the model verification can be considered as very encouraging. They demonstrate feasibility of applying this WBL-model for solving a variety of practical problems (for example, listed in Polnikov (2009)), and the possibility of further development of the model based on the regulations formulated here. Of course, this will require a significant expansion of the empirical database, especially designed and suitable in accuracy, to be used for the WBL-model verification. Problems and ways of solving these issues are considered in detail in (Polnikov 2009).

**Acknowledgements**

This work was supported by RFBR, grant № 09-05-00773_a. The author is extremely grateful to Prof. Vladimir Kudryavtsev for numerous discussions and critical comments and for familiarization with the numerical codes for calculating the HFS-shape. I am also grateful to Prof. Alex Babanin for kindly providing the data of synchronous measurement for the boundary layer parameters and two-dimensional wave spectrum.